\documentclass[referee]{raa}            
\usepackage{graphicx,times}

\providecommand{\bm}[1]{\mbox{\boldmath$#1$\unboldmath}}

\begin{document}

   \title{ Transformation between $\tau$ and TCB for Deep Space
   Missions under IAU Resolutions
 $^*$
\footnotetext{\small $*$ Supported by the National Natural Science
Foundation of China.}
}

 \volnopage{ {\bf 0} Vol.\ {\bf 0} No. {\bf 00}, 000--000}
   \setcounter{page}{1}

 \author{Xue-Mei Deng
      \inst{1}
   }

   \institute{Purple Mountain Observatory, Chinese Academy of Sciences,
   Nanjing 210008, China; \email{xmd@pmo.ac.cn}\\
   \vs \no
   {\small Received [year] [month] [day]; accepted [year] [month] [day] }
}

\abstract{ For tracking a spacecraft and doing radio science, the
transformation between the proper time $\tau$ given by a clock
carried on board a spacecraft and the barycentric coordinate time
(TCB) is investigated under IAU resolutions. In order to show more
clearly physical pictures and improve computational efficiency, an
analytic approach is adopted. After numerical checks, it shows this
method is qualified for a Mars orbiter during one year, especially being good at describing the
influence from perturbing bodies. Further analyses demonstrate that
there are two main effects in the transformation: the gravitational
field of the Sun and the velocity of the spacecraft in the
barycentric coordinate reference system (BCRS). The whole
contribution of them is at the level of a few sub-seconds.
 }

   \authorrunning{X.-M. Deng}            
   \titlerunning{Transformation between $\tau$ and TCB for Deep Space
   Missions under IAU Resolutions }  
   \maketitle


%
%
\section{Introduction}           

Last a few decades see the enormous improvements of the accuracy of
measurements and the unprecedented progress in techniques. It makes
the general relativity (GR) become an inevitable part of the data
processing in the high-precision observations. Thus, the first order
post-Newtonian (1PN) general relativistic theory of astronomical
reference frames based on \cite{brum89} and
\cite{dam91}, was adopted by General Assembly of
the International Astronomical Union (IAU) in 2000 (\cite{sof03}).
Likewise, GR plays an important role for deep space missions in
navigation and scientific experiments.

For example, the radio link connecting a spacecraft and a ground station has
been a sensitive and useful tool for probing the interior structure of a body
in the Solar System. Some signals from these intriguing but subtle effects
might entangle with those due to the curved spacetime.
This work is then motivated as the first step to construct an
applicable and consistent relativistic framework that will be able
to separate the planetary information from GR ``bias''. On
the other hand, the radio link in the interplanetary space
could test theories of gravity. In 2003, the Cassini spacecraft had
confirmed GR to an accuracy of $10^{-5}$ by
Doppler tracking in the spacecraft solar conjunction (\cite{ber03}). This
result has not only verified GR
but also ruled out some theories which unsatisfied the corresponding
condition. Deep space missions might also be opening a new window to
some new physical laws at the scale of the solar system with
the challenge of unexplained anomalies (\cite{and98,and08}).
This is the reason why a complete data reduction framework should be
established robustly in the first place to interpret the
observation data.

A general scheme for data reduction based on a relativistic
framework is represented as follows. Starting from a Lagrangian based theory
of gravity, the metric of the Solar System can be obtained by the post-Newtonian approximation (\cite{cha65}).
A global reference system covering the region of the
whole spacetime is introduced to describe the orbital motions of the bodies
in the Solar System. Some local reference systems are also introduced and each of them covers the nearby region
of a body to define the multiple moments of the body and describe the motions of its massless satellites.
However, most of current data reductions, including lunar
laser ranging, are conducted in the global frame. Thus, it involves the
coordinate transformation between the global
frame and the local one. This transformation has been intensively studied
by \cite{brum89}, \cite{dam91},
\cite{kli00},\cite{kop04} and \cite{xie10}. Within this
relativistic framework, the motions of spacecrafts, celestial bodies, light rays (photons) and
observers in the Solar System would be adequately represented in different reference frames.
The task is to make a relativistic model for a specific kind of observations with some physical or conventional quantities.

In the above process, different time scales exist within the
relativistic framework by a contrast to Newton's idea of absolute spacetime.
A clock on board a spacecraft gives the proper
time $\tau$, which is a physical time. To deal with the propagation of the  signals emitted by a spacecraft in the Solar System, the
Barycentric Celestial Reference System (BCRS) is usually used. It has a coordinate time component, called the barycentric coordinate time
(TCB). Therefore $\tau$ needs to be connected with TCB at first for the whole radio link. This is one of
our motivation in the research work. In general, a numerical method
or an analytic one could be adopted for discussing this
transformation. Although the numerical method is more competent for
computing, inverting and predicting astronomical events and
phenomena, it is not enough to provide some physical information.
With a practical case in hand, the method can not distinguish the
leading terms, the secular terms accumulated with time and the
negligible terms from the numerical results. Besides, the
presence of hundreds of terms with using higher approximation (for
example, from 1PN to 2PN) makes these problems more complicated.
However, the analytic method is extraordinarily good at these.
Especially, the computational process by the analytic method is more
time-saving and efficient. In the gauge-invariant point of view, some
spurious coordinate-dependent effects can be removed by the analytic
method. Thus a more efficient and unambiguous method should be
found for providing to the above advantages. This is the another
motivation of this paper.

To sum up, as a first step, employed an analytic method, this
work mainly focuses on the transformation between the proper
time $\tau$ on the spacecraft and TCB under IAU resolutions.
It shows there exists the difference of
two time system between on the spacecraft and on the global system.
This transformation will be applied for connecting the emitted signal
with the light propagation. It will also be applied by tracking, telemetry
and control in ground stations.

We summarize some conventions and notations used in the paper. The
metric signature is $(-,+,+,+)$; $G$ is the Newtonian constant of
gravitation; $c$ is the velocity of the light and
$\epsilon\equiv1/c$; The capital subscripts $A$, $B$, $C$ $\ldots$
refer to the gravitating bodies in the solar system; The subscripts
$T$ and $s$ denote respectively quantities related to the target
body and the spacecraft; The Latin indices $i,j,k\ldots$ denote
three-dimensional space components; The symmetric and trace-free
(STF) part of a tensor $I^{ij}$ is denoted by $I^{<ij>}$; We also
use multi-index notations such as $I^{<L>}\equiv
I^{<i_{1}i_{2}\ldots i_{l}>}$. Section 2 is devoted to an analytic
expression for the transformation between $\tau$ and TCB under IAU
resolutions. Considering a Mars mission, the comparison between the
numerical method and our analytic one is described in Section 3.
Then, in Section 4, some results are derived with our analytic
method. Finally, the conclusion and discussion are outlined in
Section 5.

\section{Model and analytic expression}

In BCRS, the metric
tensor under IAU resolutions (\cite{sof03}) reads as
\begin{eqnarray}
\label{g00}
g_{00}&=&-1+\epsilon^{2}2w-\epsilon^{4}2w^{2}+\mathcal{O}(5),\\
\label{g0i}
g_{0i}&=&-\epsilon^{3}4w^{i}+\mathcal{O}(5),\\
\label{gij}
g_{ij}&=&\delta_{ij}(1+\epsilon^{2}2w)+\mathcal{O}(4),
\end{eqnarray}
where $w$ and $w^{i}$ are respectively scalar and vector potentials.
And $\mathcal{O}(n)$ means of the order $\epsilon^{n}$. Then, the
transformation between the proper time of a spacecraft $\tau_{s}$
and TCB ($t$) can be done by integrating the following equations
\begin{eqnarray}
\label{tau}
\frac{\mathrm{d}\tau_{s}}{dt}&=&1-\epsilon^{2}\bigg(w+\frac{1}{2}v^{2}_{s}\bigg)
+\epsilon^{4}\bigg(\frac{1}{2}w^{2}+4w^{k}v^{k}_{s}-\frac{3}{2}wv^{2}_{s}-\frac{1}{8}v^{4}_{s}\bigg)+\mathcal{O}(5).
\end{eqnarray}
In principle, $w$ and $w^{i}$ should be expressed as the local
multipole moments. But, if
we only consider $N$-point masses with spins, Eq. (\ref{tau}) yields
\begin{eqnarray}
\label{tau1}
\frac{\mathrm{d}\tau_{s}}{dt}&=&1-\epsilon^{2}\bigg(\sum_{A}\frac{Gm_{A}}{r_{sA}}+\frac{1}{2}v^{2}_{s}\bigg)\nonumber\\
&&-\epsilon^{4}\bigg[\frac{1}{8}v^{4}_{s}-\frac{1}{2}\sum_{A}\frac{G^{2}m^{2}_{A}}{r^{2}_{sA}}
+2\sum_{A}\frac{Gm_{A}}{r_{sA}}v^{2}_{A}+\frac{3}{2}\sum_{A}\frac{Gm_{A}}{r_{sA}}v^{2}_{s}\nonumber\\
&&-\frac{1}{2}\sum_{A}\frac{Gm_{A}}{r^{3}_{sA}}(r^{k}_{sA}v^{k}_{A})^{2}
-4\sum_{A}\frac{Gm_{A}}{r_{sA}}v^{k}_{s}v^{k}_{A}\nonumber\\
&&-4\sum_{A}\frac{G}{r^{3}_{sA}}\varepsilon^{k}_{~ij}S^{i}_{A}r^{j}_{sA}v^{k}_{s}
-\frac{1}{2}\sum_{A}\sum_{B\neq
A}\frac{G^{2}m_{A}m_{B}}{r_{sA}r_{sB}}\nonumber\\
&&-\sum_{A}\sum_{B\neq
A}\frac{G^{2}m_{A}m_{B}}{r_{sA}r_{AB}}+\frac{1}{2}\sum_{A}\sum_{B\neq
A}\frac{G^{2}m_{A}m_{B}}{r_{sA}r^{3}_{AB}}r^{k}_{sA}r^{k}_{AB}\bigg],
\end{eqnarray}
where 1PN masses are obtained by using the method of two effective
time-dependent masses of the $A$-th body
\begin{eqnarray}
\tilde{\mu}_{A}&=&m_{A}\bigg\{1+\epsilon^{2}\bigg[-\sum_{B\neq
A}\frac{Gm_{B}}{r_{AB}}+\frac{3}{2}v^{2}_{A}\bigg]\bigg\}+\mathcal{O}(4),\\
\mu_{A}&=&m_{A}\bigg\{1+\epsilon^{2}\bigg[-\sum_{B\neq
A}\frac{Gm_{B}}{r_{AB}}+\frac{1}{2}v^{2}_{A}\bigg]\bigg\}+\mathcal{O}(4),
\end{eqnarray}
based on \cite{bla98}. And $r_{sA}=|\bm{x}_{s}-\bm{x}_{A}|$,
$r_{AB}=|\bm{x}_{A}-\bm{x}_{B}|$. $\bm{x}_{s}$ and
$\bm{x}_{A}$ respectively denote positions of the spacecraft and
the $A$-th body in BCRS. $\bm{v}_{s}$ and $\bm{v}_{A}$
respectively denote velocities of the spacecraft and the $A$-th body
in BCRS. $\varepsilon^{k}_{~ij}$ is the fully antisymmetric
Levi-Civita symbol and $S^{i}_{A}$ is the spin of the $A$-th body.
In this paper, we mainly consider the effects for terms of Eq.
(\ref{tau1}) in the order of $\epsilon^{2}$ on the transformation.
Namely,
\begin{eqnarray}
\label{tau2}
\frac{\mathrm{d}\tau_{s}}{dt}&=&1-\epsilon^{2}\bigg(\sum_{A}\frac{Gm_{A}}{r_{sA}}+\frac{1}{2}v^{2}_{s}\bigg)+\mathcal{O}(\epsilon^{4}).
\end{eqnarray}
The first term in the order of $\epsilon^{2}$ is a dynamical term
which is contributed from $N$-body's gravitational fields.
The second term in Eq. (\ref{tau2}) at the order $\epsilon^{2}$ is
a kinematic term which comes from the velocity of the spacecraft.
For the dynamical term, we split it into two parts:
\begin{eqnarray}
\sum_{A}\frac{Gm_{A}}{r_{sA}}&=&\sum_{A\neq
T}\frac{Gm_{A}}{r_{sA}}+\frac{Gm_{T}}{r_{sT}},
\end{eqnarray}
where the first one comes from the contribution of perturbing bodies and
the second one is caused by the target body for the deep space mission.
There exists a small quantity $q\equiv r_{sT}/r_{AT}$, which describes the
distance between the spacecraft and the target body divided by the
distance between the target body and the perturbing body. For the perturbing terms, they can be expanded by $q^{k}$ ($k=0,1,2,\ldots$) and show
\begin{eqnarray}
\sum_{A\neq T}\frac{Gm_{A}}{r_{sA}} &=&\sum_{A\neq
T}\frac{Gm_{A}}{|\bm{x}_{A}-\bm{x}_{T}-(\bm{x}_{s}-\bm{x}_{T})|}\nonumber\\
&=&\sum_{A\neq
T}\sum_{k=0}^{l}\frac{(2k-1)!!}{k!}\frac{Gm_{A}}{r^{2k+1}_{AT}}r^{<K>}_{AT}r^{<K>}_{sT}
+\mathcal{O}(\sum_{k=l+1}^{\infty})\nonumber\\
&=&\underbrace{\underbrace{\underbrace{\sum_{A\neq
T}\frac{Gm_{A}}{r_{AT}}}_{l=0}+\sum_{A\neq
T}\frac{Gm_{A}}{r^{3}_{AT}}r^{k}_{AT}r^{k}_{sT}}_{l=1}
+\frac{3}{2}\sum_{A\neq
T}\frac{Gm_{A}}{r^{5}_{AT}}r^{<i}_{AT}r^{j>}_{AT}r^{<i}_{sT}r^{j>}_{sT}}_{l=2}+\mathcal{O}(\sum_{k=3}^{\infty}).
\end{eqnarray}
For the dynamical term of perturbations, the above analytic
expression will be converged with the increase of index $k$. In our
research, we mainly focus on the first three terms, which correspond to
$l=2$. In the next section, we will prove the difference between the
analytic method and the numerical one for the perturbations
is negligible for current accuracy.

Our task is to give the analytic expression of Eq. (\ref{tau2}) at
the order of $\epsilon^{2}$. The positions and velocities of the
bodies and the spacecraft in the solar system are obtained by
treating them as $N$ two-body problems. For example, the motions of
eight planets with respect to the Sun are considered as 8 two-body
problems and the motion of the spacecraft with respect to its target
body is also considered as a two-body problem. For planet $A$, its
position $\bm{r}^{Heli}_{A}$ and velocity $\bm{v}^{Heli}_{A}$ are
expressed with the orbital elements in the heliocentric coordinate
system as two-body problem (\cite{mur00}). Those elements are
changing with time, such as $a_{A}=a_{A0}+\dot{a}_{A}T_{eph}$,
$e_{A}=e_{A0}+\dot{e}_{A}T_{eph}$ and so on, based on Table 1 in
Technical Report of JPL (\cite{sta}), where $T_{eph}$ is the number
of centuries past J2000.0. With the positions and velocities of
eight planets in the heliocentric coordinate system obtained, the
position and velocity of the solar system barycenter (SSB) in the
heliocentric coordinate system are then respectively obtained by
$\sum_{A}m_{A}\bm{r}^{Heli}_{A}/\sum_{A}m_{A}$ and
$\sum_{A}m_{A}\bm{v}^{Heli}_{A}/\sum_{A}m_{A}$. Using the positions
of the planets and SSB in the heliocentric coordinate system, we
could obtain the positions and the velocities of the Sun, Mercury,
Venus, the Earth-Moon Barycenter (EMB), Mars, Jupiter, Saturn,
Uranus and Neptune in BCRS. For $r^{i}_{sT}$, we solve it from the
two-body problem in the equatorial reference system of the target
body (\cite{mur00}). Furthermore, we rotate vector $r^{i}_{sT}$ from
the equatorial reference system to the International Celestial
Reference System (ICRS) based on the procedures recommended by the
IAU/IAG Working Group on cartographic coordinates and rotational
elements (\cite{USGS}). The propose of above rotations is to deal
with the coupling terms with vectors calculated in different
reference systems.

For the kinematic term of Eq. (\ref{tau2}), we only focus on the two-body interactions and omit others bodies' perturbations
\begin{eqnarray}
\label{v}
v^{2}_{s}&=&(\bm{V}_{s}+\bm{v}_{T})\cdot(\bm{V}_{s}+\bm{v}_{T})+\mathcal{O}(\epsilon^{2},\mathrm{others})\nonumber\\
&=&v^{2}_{T}+V^{2}_{s}+2\bm{v}_{T}\cdot\bm{V}_{s}+\mathcal{O}(\epsilon^{2},\mathrm{others})\nonumber\\
&=&Gm_{\odot}\bigg(\frac{2}{r_{T\odot}}-\frac{1}{a_{T}}\bigg)+Gm_{T}\bigg(\frac{2}{r_{sT}}-\frac{1}{a_{s}}\bigg)
+2\bm{v}_{T}\cdot\bm{V}_{s}+\ldots,
\end{eqnarray}
where subscripts ``$s$", ``$\odot$" and ``$T$" denote the
terms related to the spacecraft, the Sun and the target body, respectively.
And $\bm{v}_{T}$ denotes the velocity vector of the target body in BCRS, and $\bm{V}_{s}$ denotes the velocity vector of the
spacecraft in the target body's local reference system. For $\bm{v}_{T}\cdot\bm{V}_{s}$ in Eq. (\ref{v}), we must put the two vectors
in the same coordinate system such as BCRS. $\bm{V}_{s}$ can be written as $(V_{s},0,0)^{T}$ in (U, N, W) triad
where U points to the tangent direction of the orbit. Furthermore, we rotate this vector to  the (S, T, W) triad
where S points to the radial direction, then to the equatorial plane of the target body and finally to BCRS. Such a transformation is conducted by $R_{3}(-90^{\circ}-\alpha_{T})R_{1}(\delta_{T}-90^{\circ})R_{3}(-\Omega)R_{1}(-i)R_{3}(-\omega-f)R_{3}(\theta)(V_{s},0,0)^{T}$,
where $\alpha_{T}$ and $\delta_{T}$ are ICRF equatorial coordinates at epoch J2000.0 for the north pole of one target body;
$\Omega$ denotes longitude of ascending node for the spacecraft; $i$ denotes inclination of orbit for the spacecraft;
$\omega$ denotes longitude of periastron for the spacecraft; $f$ is the true anomaly and $\theta$ is the angle between
the tangent direction and transverse direction of the orbit, and $\cos\theta=(1+e\cos f)/\sqrt{1+2e\cos f+e^{2}}$.

Then, the analytic relation between $\tau_{s}$ and $t$ is
\begin{eqnarray}
\label{TA} \tau_{s}-t&=&-\epsilon^{2}\int\bigg[\sum_{A\neq
T}\frac{Gm_{A}}{r_{AT}}+\sum_{A\neq
T}\frac{Gm_{A}}{r^{3}_{AT}}r^{k}_{AT}r^{k}_{sT}
+\frac{3}{2}\sum_{A\neq
T}\frac{Gm_{A}}{r^{5}_{AT}}r^{<i}_{AT}r^{j>}_{AT}r^{<i}_{sT}r^{j>}_{sT}+\frac{Gm_{T}}{r_{sT}}\bigg]\mathrm{d}t\nonumber\\
&&-\epsilon^{2}\int\bigg[Gm_{\odot}\bigg(\frac{1}{r_{T\odot}}-\frac{1}{2a_{T}}\bigg)+Gm_{T}\bigg(\frac{1}{r_{sT}}-\frac{1}{2a_{s}}\bigg)
+\bm{v}_{T}\cdot\bm{V}_{s}\bigg]\mathrm{d}t\nonumber\\
&&+\mathcal{O}(\epsilon^{2},l\geq3,\mathrm{others}),
\end{eqnarray}
where the positions and the velocities of
$N$-body and the spacecraft can easily obtained just by two-body problem solutions. Compared to
the numerical method, this analytic approach is more efficient in computation. In next section, we will prove it is qualified by the numerical check.

\section{Numerical check}
\label{Num}

In this section, we will check our analytic result by comparison
with the numerical results under a Mars
mission. We simulate a spacecraft has a very large elliptical
orbit around Mars from Nov. 01, 2012 to Nov. 01, 2013. Its orbital
inclination to the Mars equator is about $5^{\circ}$. The apoapsis
altitude is $80,000$km and the periapsis altitude is $800$km, with
period of about 3 days.

\begin{figure}
\includegraphics[width=60mm]{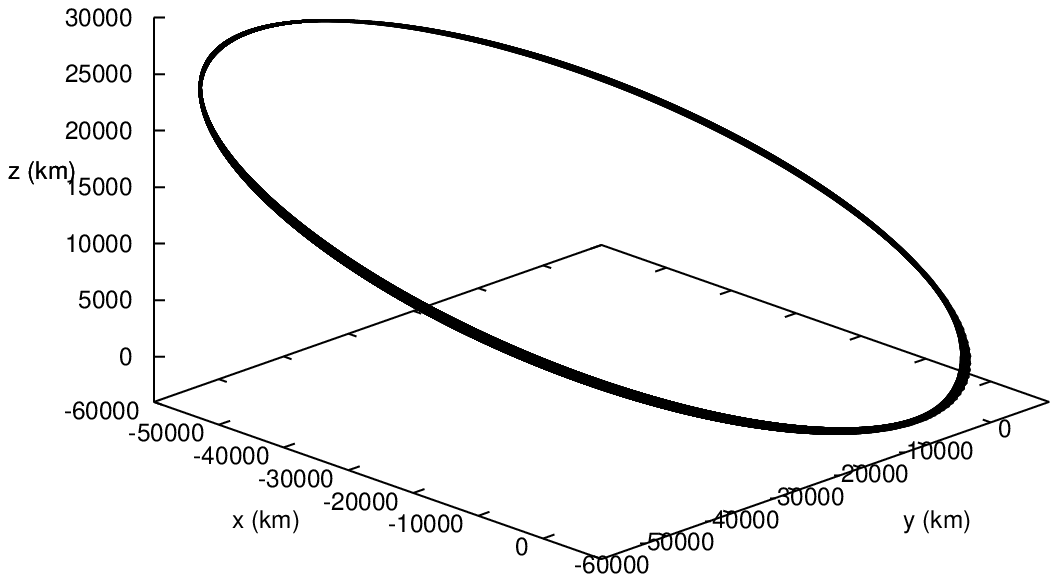}
\includegraphics[width=40mm]{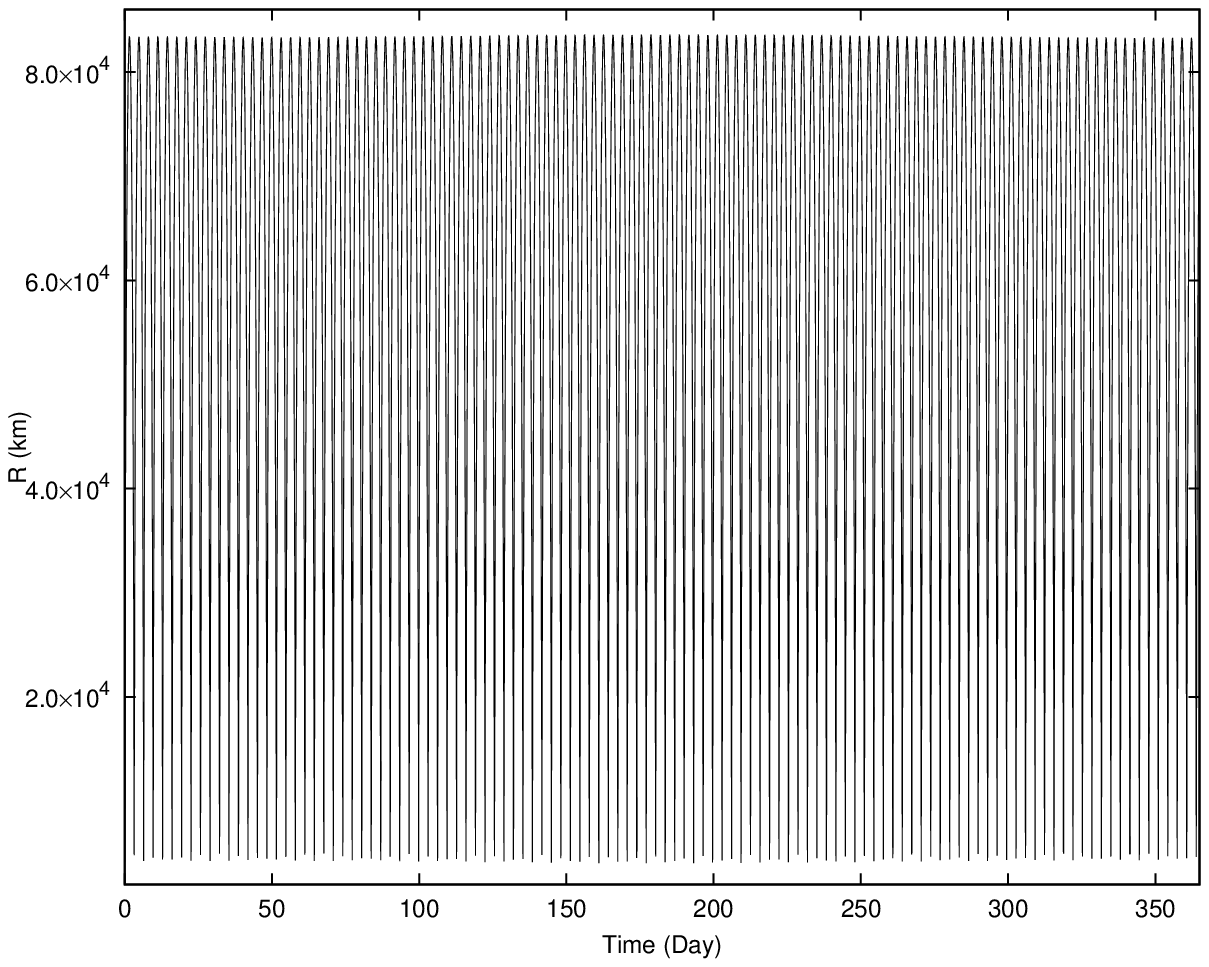}
\includegraphics[width=40mm]{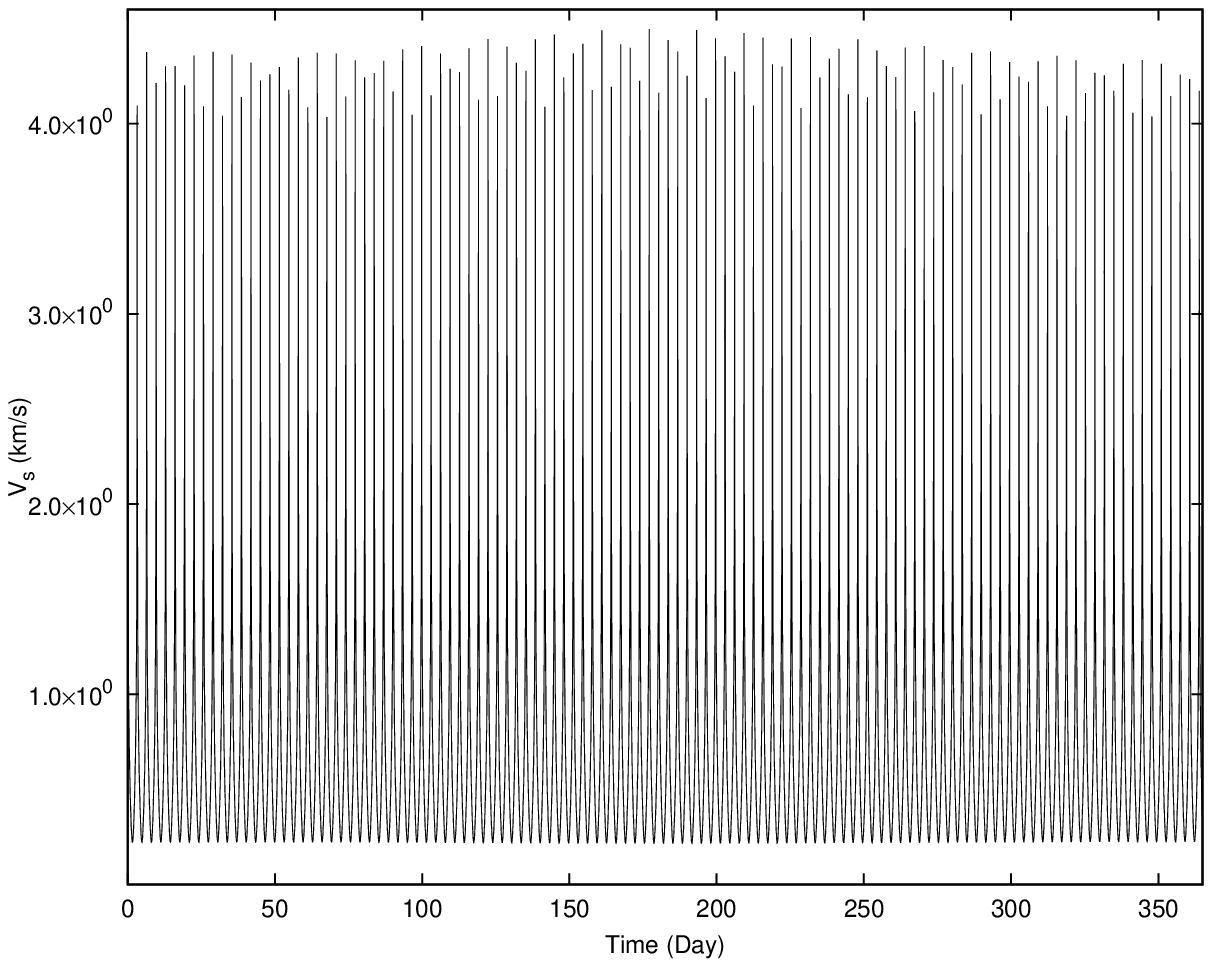}
\caption{Left: 3D orbital motion of the spacecraft for probing Mars
(in kilometers). Middle: The distance $R$ (in kilometers) between
Mars' center and the spacecraft versus the time (in days). Right:
The velocity of the spacecraft $V_{s}$ (in kilometers per second)
with respect to the center of mass of Mars versus the time (in
days). }
   \label{Fig1}
\end{figure}

In the simulation, the positions and the velocities of the planets
and the Sun are read from the ephemeris DE405. For the initial
conditions of the spacecraft, we calculate them from its orbital
elements in the Mars equatorial reference frame and transfer them
into the ICRS for numerical integration. For one year mission, the
precession and nutation are negligible in the rotation elements of
Mars for this transformation. So we only consider the fixed term for
the north pole of Mars, namely, $\alpha_{0}=317^{\circ}.68143$ and
$\delta_{0}=52^{\circ}.88650$ (see \cite{USGS}). The integrator we
use is the RKF7(8) (\cite{feh68}) with fixed step-size $30$ minutes.

In Fig. \ref{Fig1}, our numerical results for the spacecraft are
displayed. The left one shows its 3D orbit and we can see a large
ellipse. The middle one shows the change of its distance from Mars
with time. We can see the max value and the min value of $R$. The
right one shows its velocity with respect to the center of mass of
Mars. With the position $\bm{x}_{s}$ and velocity $\bm{v}_{s}$ of
the spacecraft in BCRS, we can numerically calculate
\begin{eqnarray}
\label{TN}
\tau_{s}-t&=&-\epsilon^{2}\int\bigg(\sum_{A}\frac{Gm_{A}}{r_{sA}}+\frac{1}{2}v^{2}_{s}\bigg)\mathrm{d}t.
\end{eqnarray}
Since obtained from DE405, the positions of the planets depend on the
coordinate time of the planetary ephemeris: Barycentric Dynamical
Time (TDB). The relationship between TDB and TCB is
$\mathrm{TDB}=(1-L_{B})\mathrm{TCB}$ with
$L_{B}=1.550519768\times10^{-8}$ according to \cite{iau2006}.
However, this influence of $L_B$ could be negligible because it
couples with $\epsilon^{2}$.

\begin{figure}[htbp]
\begin{center}
\includegraphics[height=45mm,angle=0]{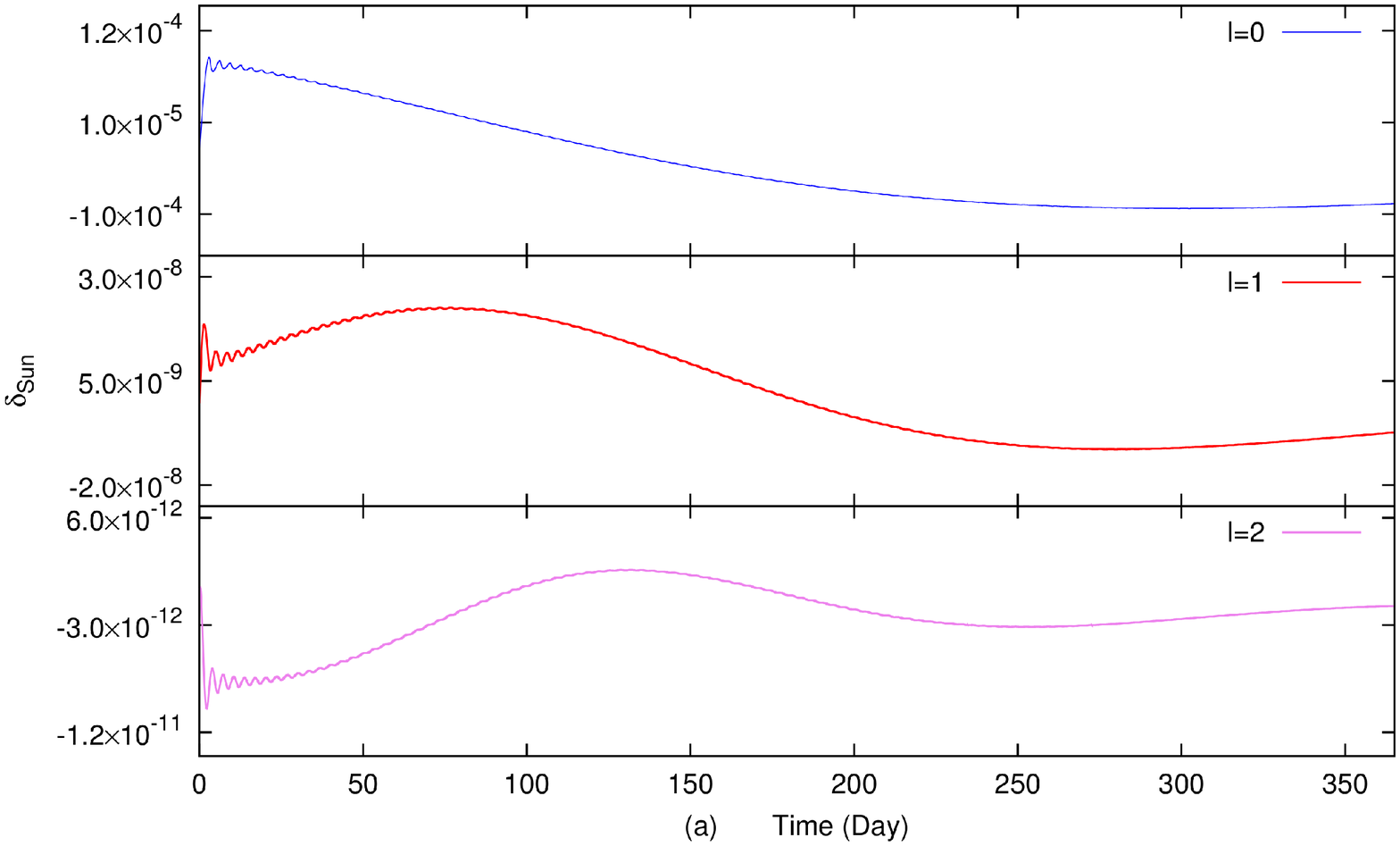}
\includegraphics[height=45mm,angle=0]{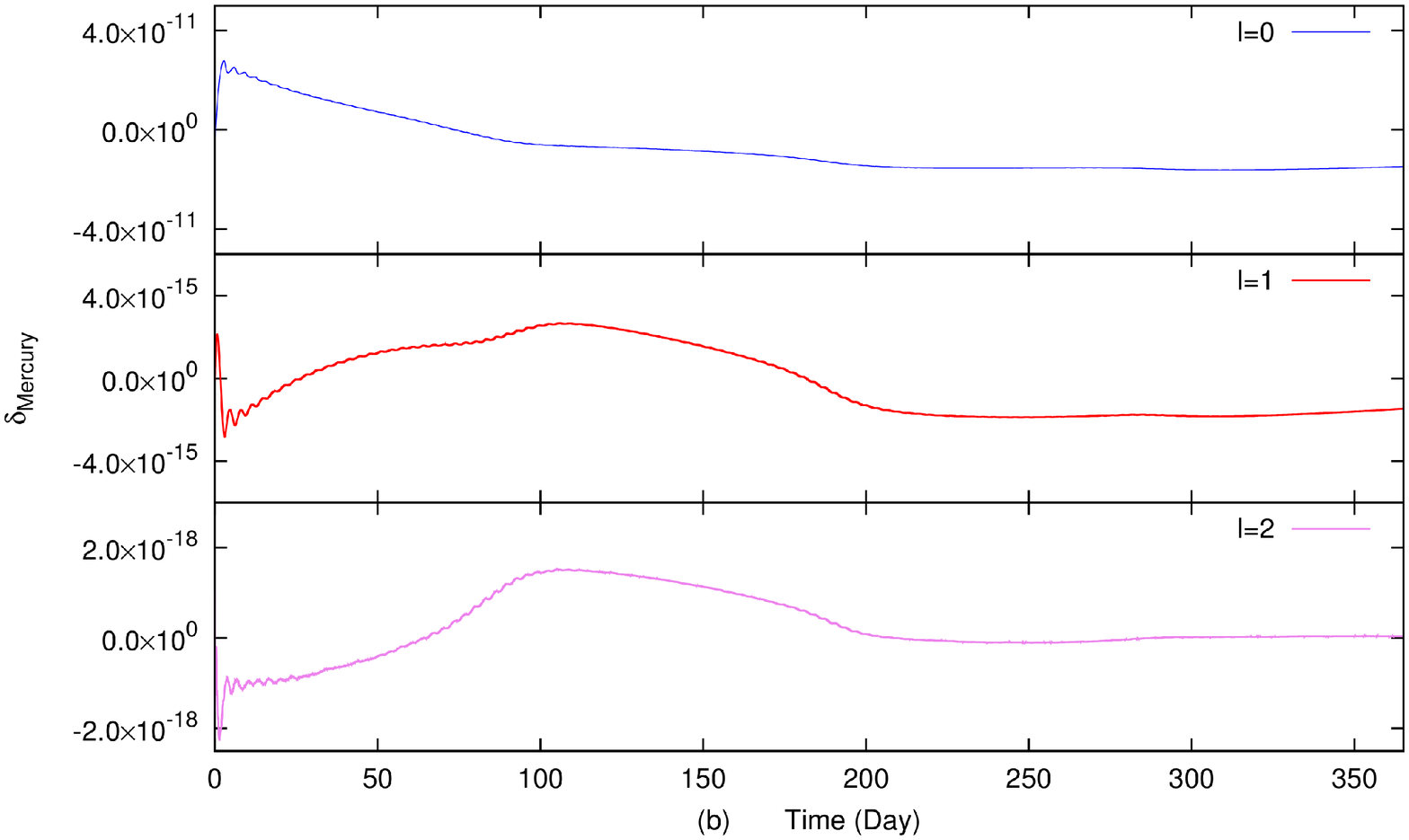}
\includegraphics[height=45mm,angle=0]{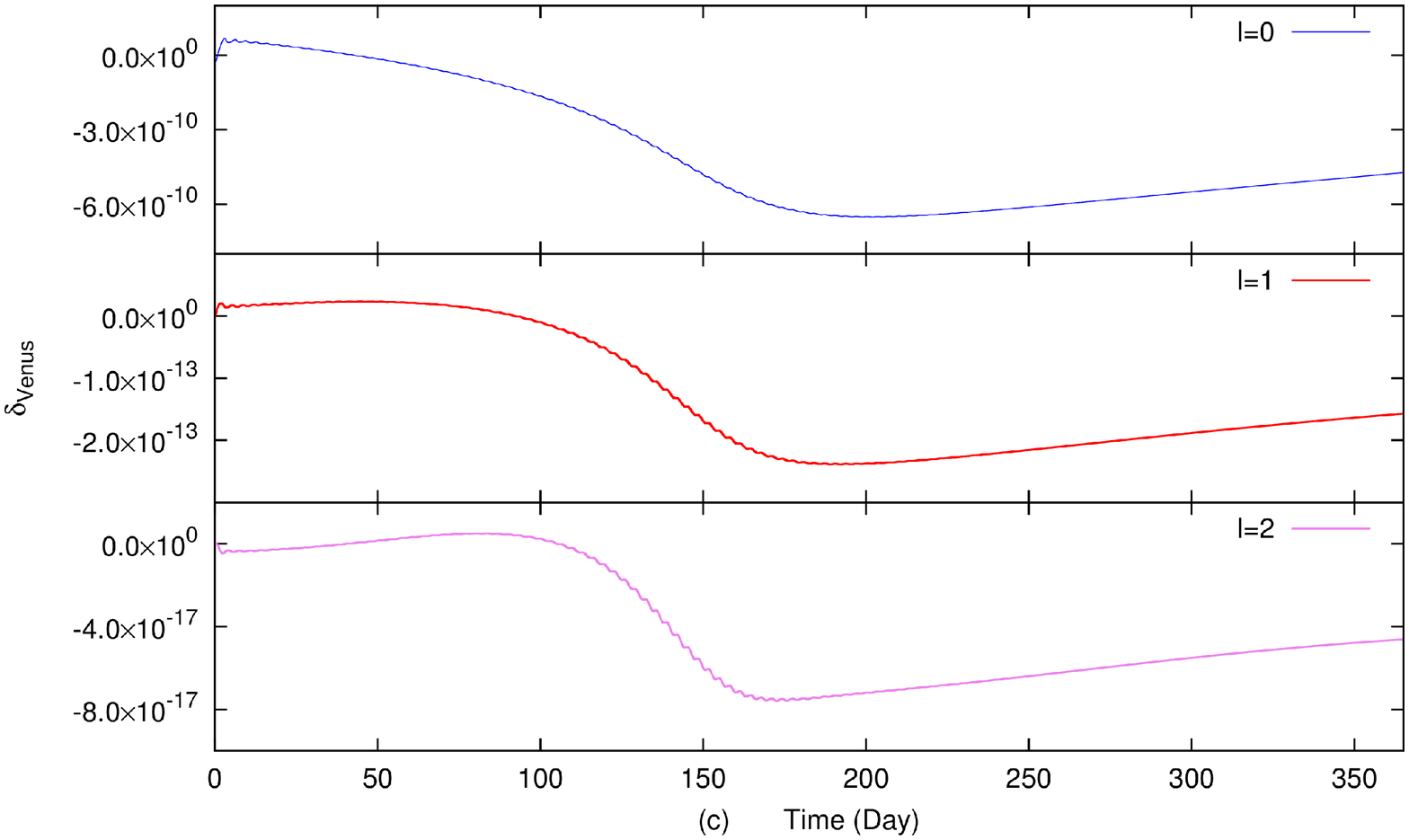}
\includegraphics[height=45mm,angle=0]{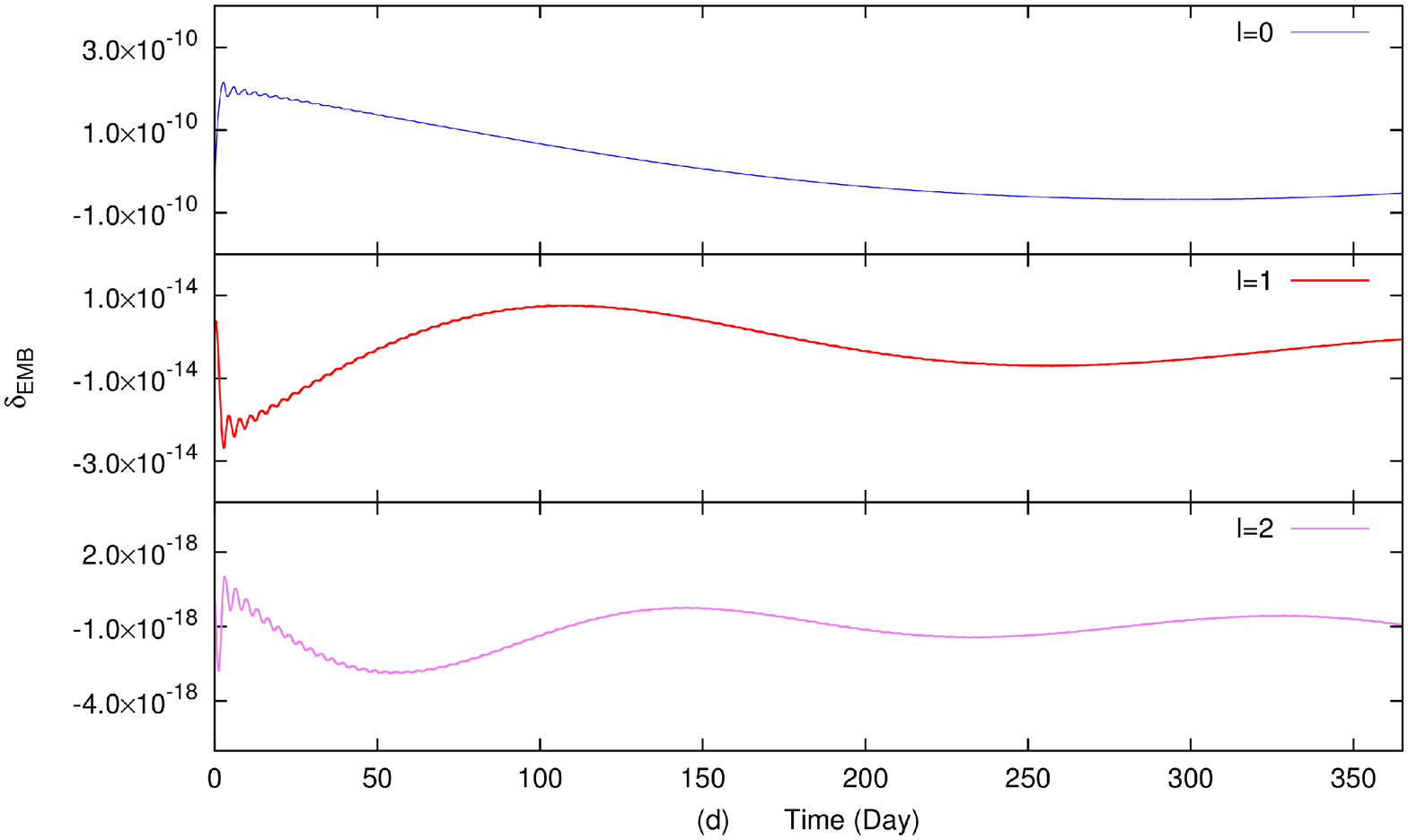}
\includegraphics[height=45mm,angle=0]{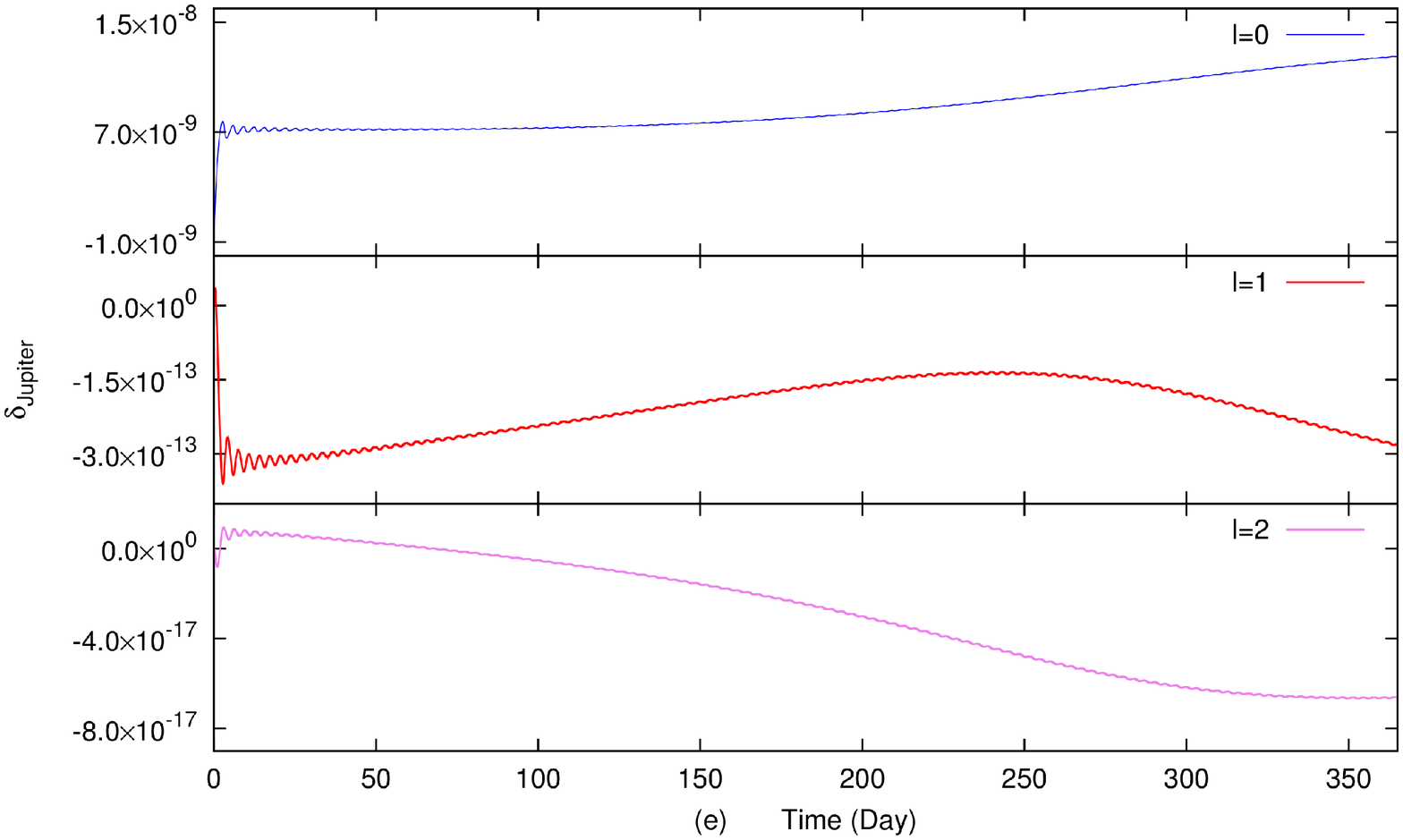}
\includegraphics[height=45mm,angle=0]{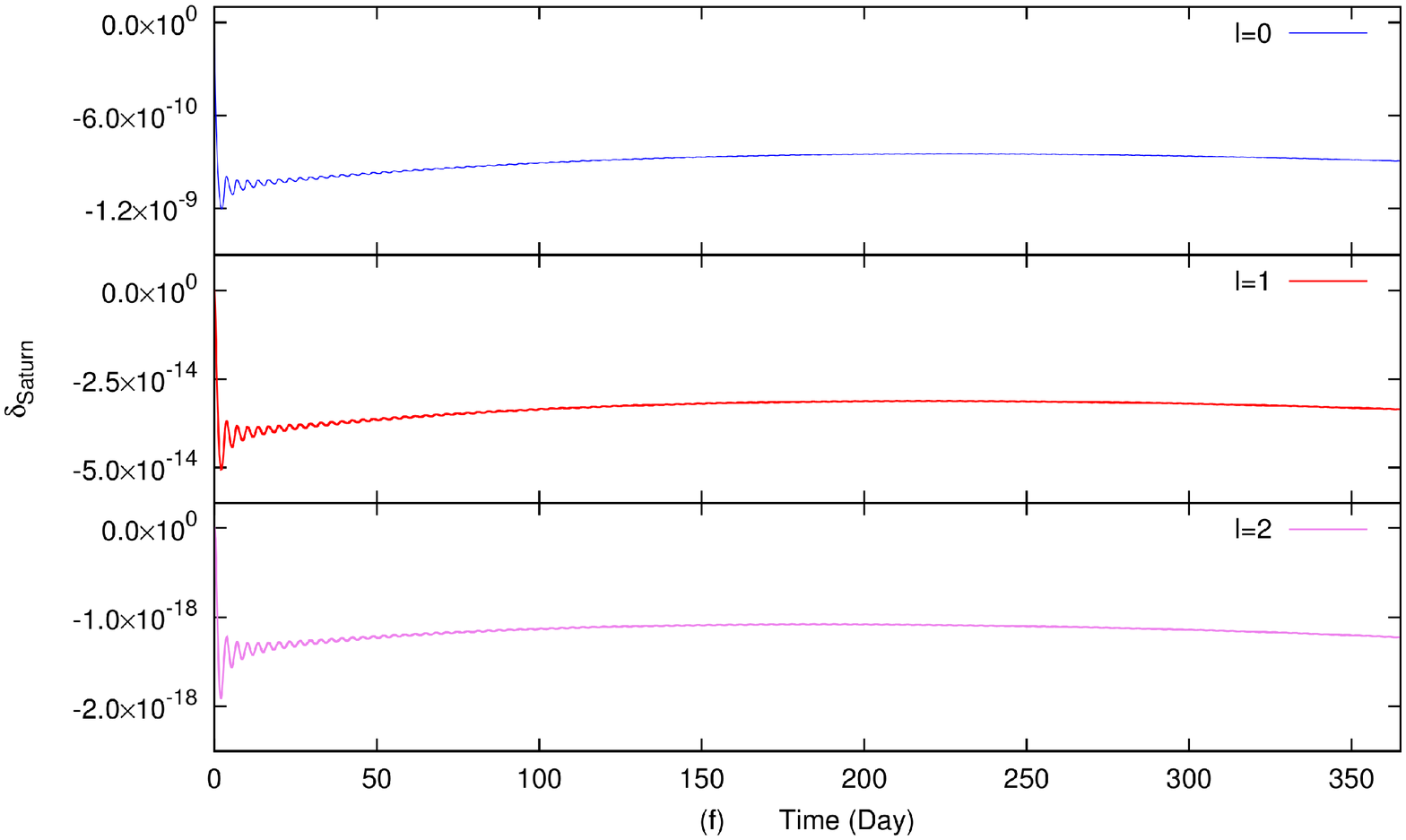}
\includegraphics[height=45mm,angle=0]{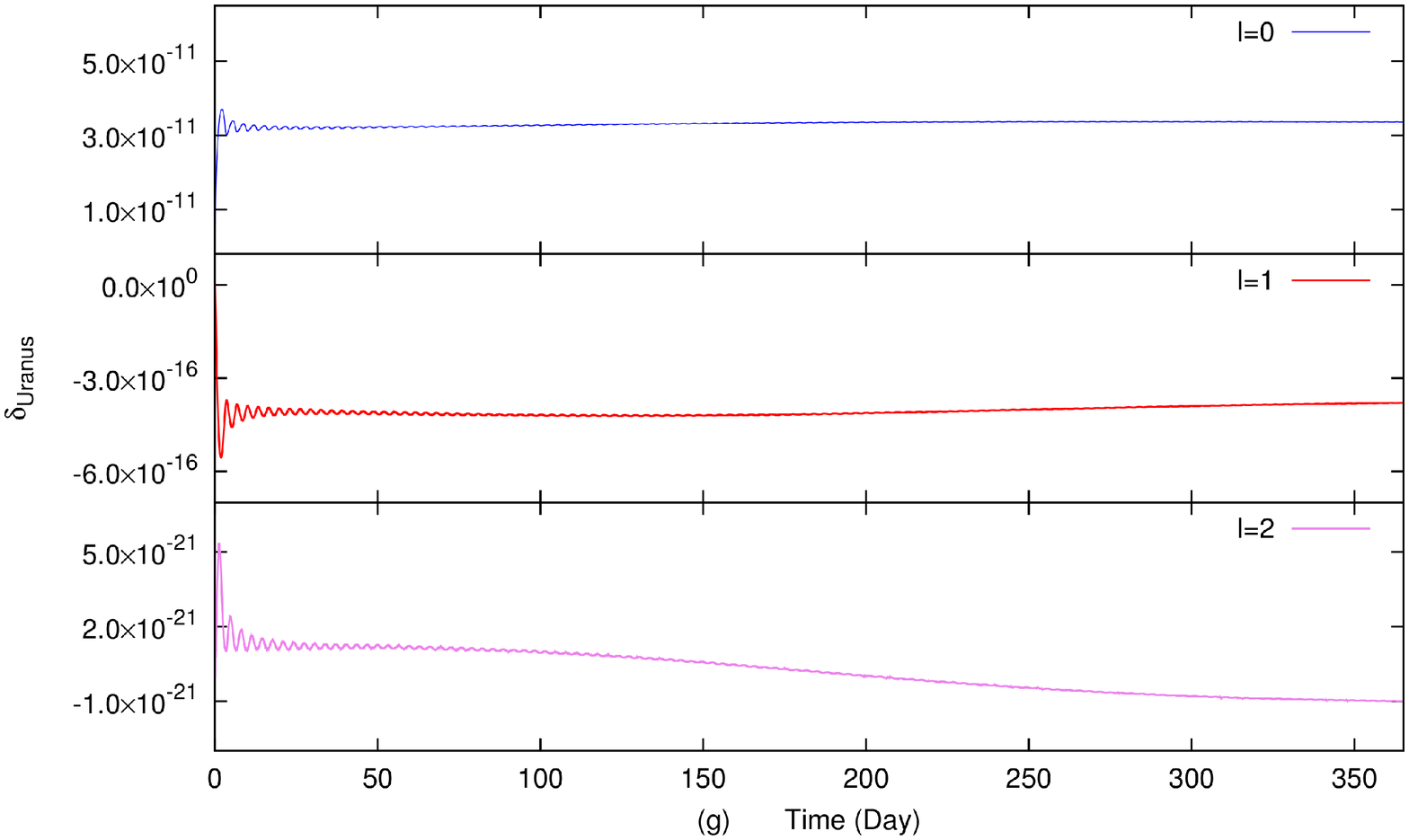}
\includegraphics[height=45mm,angle=0]{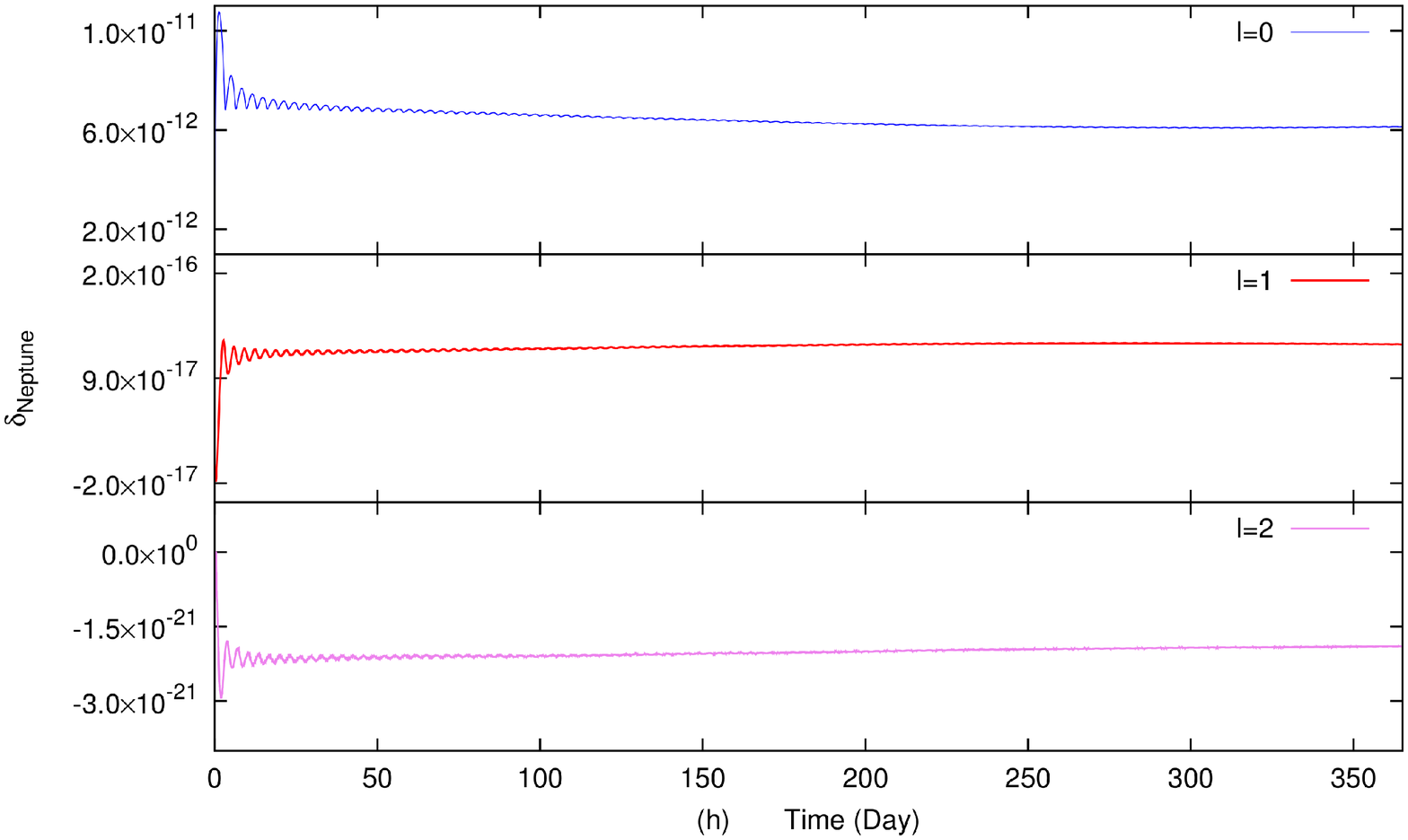}
\caption{ The normalized relative deviation between the analytic and
numerical results $\delta_{A}$ for perturbing bodies with $l=0$
(blue),$l=1$ (red),$l=2$ (violet) versus the integral time (in
days). (a) for the Sun; (b) for Mercury; (c) for Venus; (d) for the
EMB; (e) for Jupiter; (f) for Saturn; (g) for Uranus; (h) for
Neptune.}
   \label{Fig2}
   \end{center}
\end{figure}

\begin{figure}
\includegraphics[width=70mm]{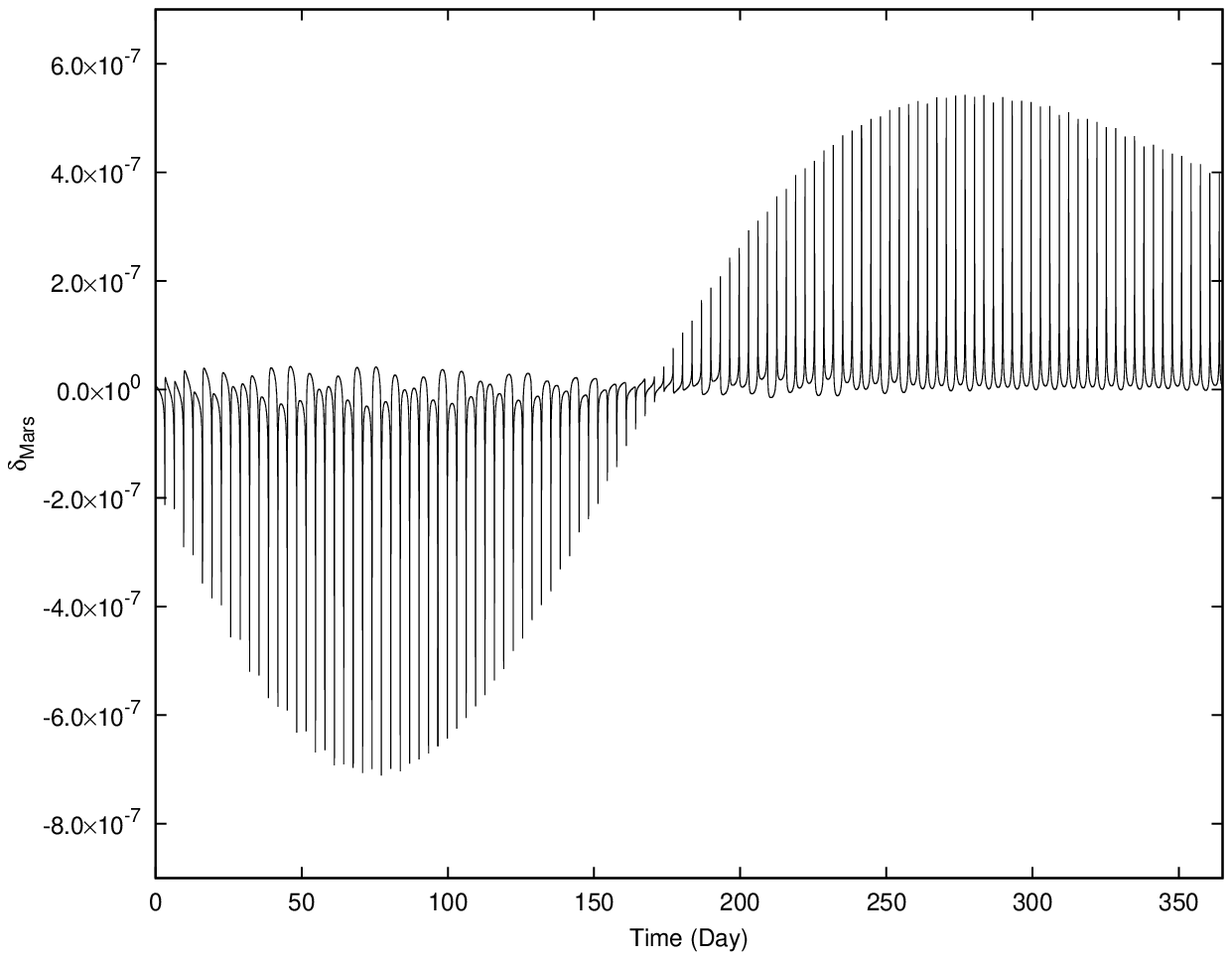}
\includegraphics[width=70mm]{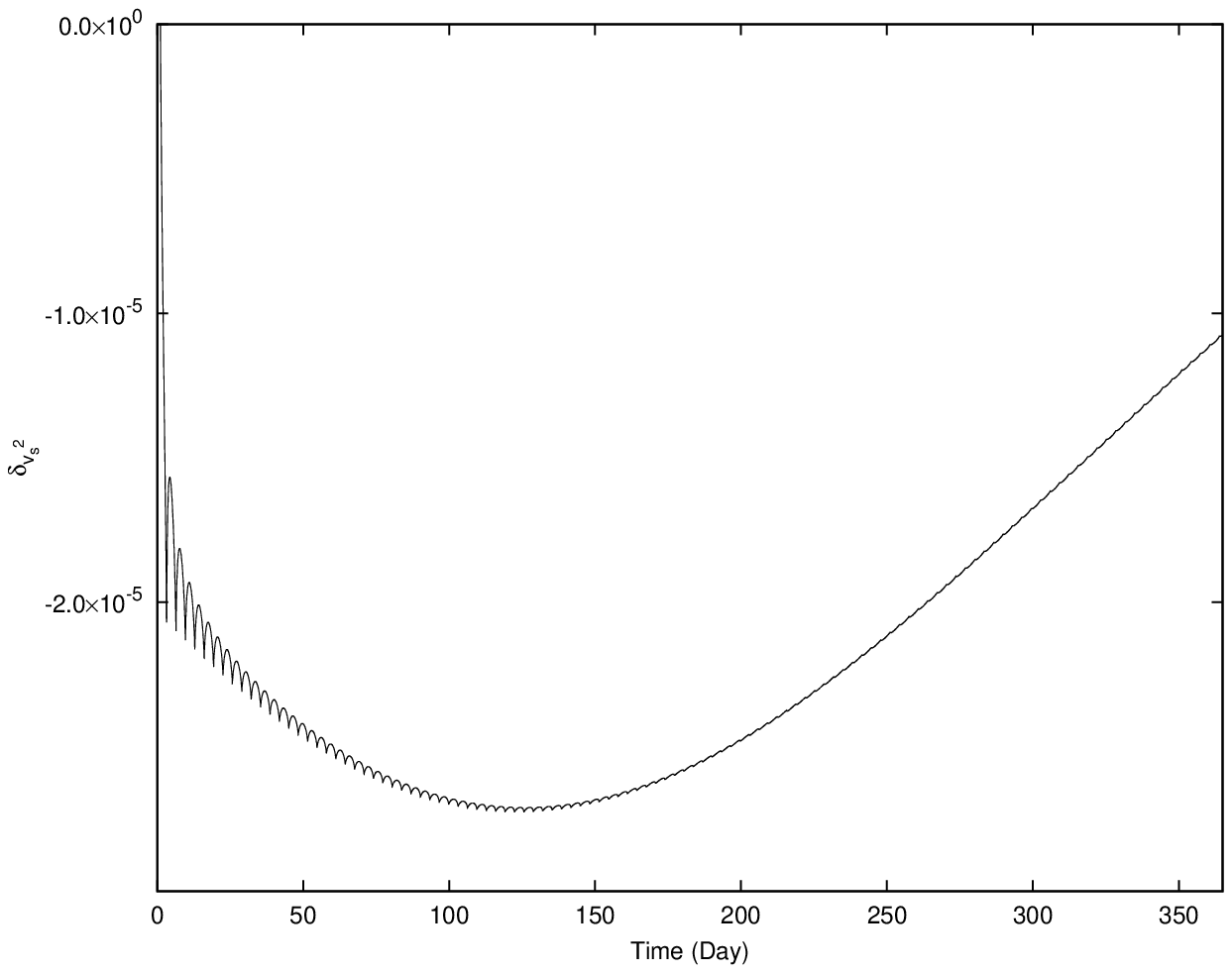}
\caption{Left:The normalized relative deviation for Mars versus the
integral time (in days). Right: The normalized relative deviation
for the velocity of the spacecraft in the BCRS versus the integral
time (in days). }
   \label{Fig3}
\end{figure}
With these numerical results, we can check our analytic approach.
Firstly, we consider the effects of the dynamical term. For
perturbations, there are three terms in the analytic expression (see
Eq. (\ref{TA})). We introduce a dimensionless quantity $\delta_{A}$
for contribution A in $\tau_{s}-t$, which is defined
as $\delta_{A}\equiv[\mathrm{analytic~ (A})-\mathrm{numerical
~(A})] / [\mathrm{numerical} ~(\tau_{s}-t)]$.
Fig. \ref{Fig2} shows
$\delta_{A}$ of the Sun, Mercury, Venus, EMB, Jupiter, Saturn,
Uranus and Neptune for $l=0$, $l=1$ and $l=2$. The contributions of
perturbations are very well described by our analytic approach
because $\delta_{A}$ decreases to $\sim 10^{-12}$ or below with
$l=2$. Although the curves of Fig. \ref{Fig2} have some
fluctuation in the beginning, they tend to be smooth with time.
For the effect of Mars in the dynamical term, the left one of
Fig. \ref{Fig3} displays the comparison between the numerical and
the analytic results. The maximum $\delta_{\mathrm{Mars}}$ is about
$10^{-7}$. The right one of Fig. \ref{Fig3} shows the numerical
check of the kinematic term and $\delta_{v_s^2}$ is about $10^{-5}$.
Both of them are caused by the fact that pure two-body problem
solutions are adopted in our analytic approach but it is full N-body
integration in the numerical simulation.

\section{Analytic results}

\begin{figure}[htbp]
\begin{center}
\includegraphics[height=70mm,angle=0]{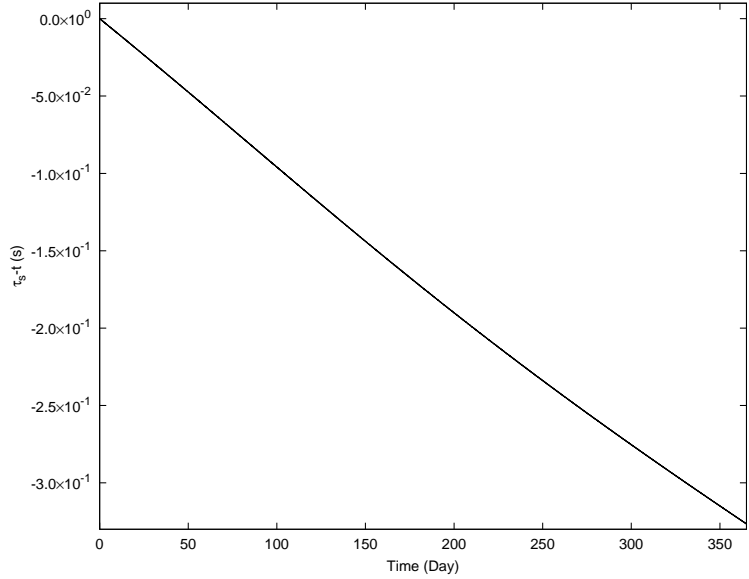}
\caption{This figure shows the difference in $\tau_{s}-t$ with the
time. }
   \label{Fig4}
   \end{center}
\end{figure}

\begin{table}[h!!!]
\small \centering
\begin{minipage}[]{120mm}
\caption[]{ Components in $\tau_{s}-t$}\label{Table 1}\end{minipage}
\tabcolsep 0.1mm
 \begin{tabular}{c|cc|c|cc|c|cc}
  \hline
   & Max (s) & Order    &  & Max (s) & Order &  & Max (s) & Order  \\ 
 \hline
Sun &0.2 & $c^{-2}$ & Mercury & $4\times10^{-8}$ & $c^{-2}$ &
EMB & $4\times10^{-7}$ & $c^{-2}$  \\
Sun  & $7\times10^{-10}$ & $c^{-4}$ & Venus & $6\times10^{-7}$ &
$c^{-2}$ &
Mars & $3\times10^{-4}$ & $c^{-2}$ \\ 
Jupiter & $7\times10^{-5}$ & $c^{-2}$ & Uranus &
$7\times10^{-7}$ & $c^{-2}$ &
$\int\frac{1}{2}v^{2}_{s}\mathrm{d}t$ & 0.1 & $c^{-2}$  \\
Saturn  & $8\times10^{-6}$ & $c^{-2}$ & Neptune & $5\times10^{-7}$ &
$c^{-2}$ &
$\int\frac{1}{8}v^{4}_{s}\mathrm{d}t$ & $2\times10^{-10}$ & $c^{-4}$ \\ 
\hline
\end{tabular}
\end{table}

\begin{figure}[htbp]
\begin{center}
\includegraphics[height=90mm,angle=0]{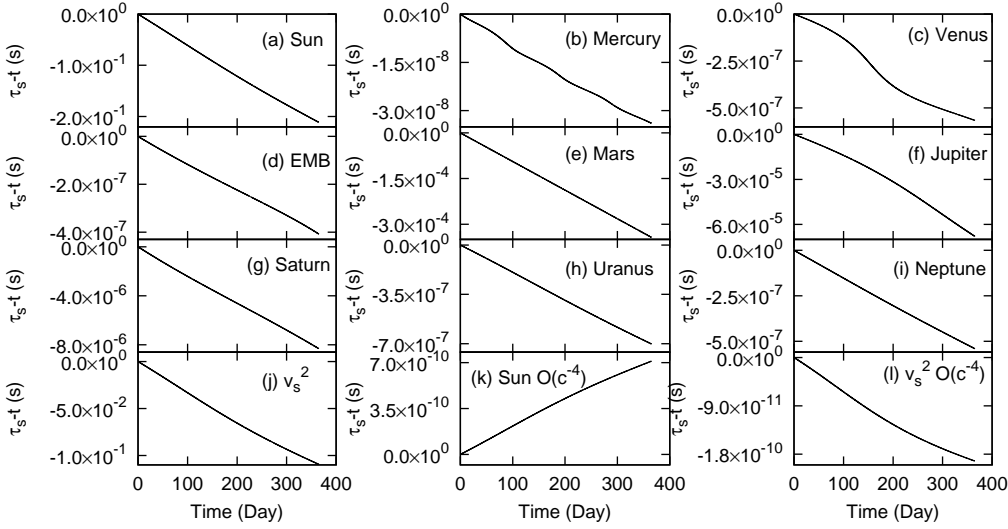}
\caption{Different terms in Eq. (\ref{TA}) versus the time. Figures
(a)-(j) denote the contributions from the Sun, Mercury, Venus, the
EMB, Mars, Jupiter, Saturn, Uranus, Neptune and $v^{2}_{s}$ at the
order of $c^{-2}$, respectively. Figures (k) and (l) denote the
effects of the Sun and the velocity of the spacecraft in
$\tau_{s}-t$ at the order of $c^{-4}$.}
   \label{Fig5}
   \end{center}
\end{figure}

Some results are derived with our analytic method after qualified by
the numerical check. Fig. \ref{Fig4} shows the curve of $\tau_{s}-t$
by Eq. (\ref{TA}). We can see the difference between the proper time
and TCB could reach the level of sub-second. This effect has two
main components: the Sun's gravitational field and the velocity of
the spacecraft in the BCRS. Fig.\ref{Fig5}(a)-(j) display the
contributions of the Sun, Mercury, Venus, the EMB, Mars, Jupiter,
Saturn, Uranus, Neptune and the velocity of the spacecraft (i.e.
$-\epsilon^{2}\int v^{2}_{s}/2\mathrm{d}t$). Since the Sun's
gravitational field and the velocity of the spacecraft in BCRS
dominate, we further consider these contribution in the next order,
namely, $\epsilon^{4}\int
G^{2}m^{2}_{\odot}/(2r^{2}_{s\odot})\mathrm{d}t$ and
$-\epsilon^{4}\int v^{4}_{s}/8\mathrm{d}t$ (see Fig.\ref{Fig3}(k)
and Fig.\ref{Fig3}(l)). These two terms in the order of
$\epsilon^{4}$ are very small around $\sim 10^{-10}$s.

Table \ref{Table 1} gives the maximum values of different effects in
the $\tau_{s}-t$. At the order of $\epsilon^{2}$, the Sun's
gravitational field and the velocity for the spacecraft have the
contributions up to a few sub-seconds, while others belong to
microsecond-level or below. At the order of $\epsilon^{4}$, the
maximum contributions of the Sun's gravitational field and the
velocity for the spacecraft are at the level of $0.1$ nanosecond. It
means if we take $1$ nanosecond as the precision of time system, the
transformation between the proper time on the spacecraft and TCB
needs to include the terms at the order of $\epsilon^{2}$ only.

If we take YingHuo-1 Mission as a technical example for Chinese
future Mars explorations, the supposed spacecraft will be equipped
with a clock such as the Ultra-Stable-Oscillator (USO),
whose instability is less than $1\times10^{-12}$ or
$2\times10^{-12}$ from 0.1 to 1000 seconds (\cite{pin09}). The
accuracy control must be done for a clock carried on board because
its accuracy will be drift as a result of various reasons. Thus, it
is almost impossible to estimate the timing error of a clock after
one year through its stability or accuracy number. And we only
discuss a time span in one year mission such as one month.
$\tau_{s}-t$ can get the level of $10^{-2}$s in one month. At the
level of microsecond $10^{-6}$s of the time accuracy, although
$\delta_{\mathrm{Mars}}$ and $\delta_{v_s^2}$ could reach $10^{-7}$
and $10^{-5}$ maximally, their maximum contributions in the
deviation of $\tau_{s}-t$ are respectively $10^{-9}$s and $10^{-7}$s
for one month, both of them less than $10^{-6}$s. It shows our
analytical approach is qualified for a Mars orbiter.

\section{Conclusion}

In this paper, the transformation between the proper time on the
spacecraft and TCB is derived under IAU resolutions.
In order to obtain more clearly
physical pictures and improve
computational efficiency, an analytic approach is employed.
A numerical simulation of a Mars mission is conducted and shows this
approach is qualified, especially being good at dealing with perturbations.
It shows that the difference between the proper time on the spacecraft and
TCB reaches the level of sub-second. And the main contributions of
this transformation come from the Sun's gravitational field and the
velocity of the spacecraft in the BCRS.

In this work, we only take two-body problem solutions, which makes the relative deviations of Mars' gravitational field
and the velocity of the spacecraft reach respectively about $10^{-7}$ and $10^{-5}$. Our
next move is to include the effect of the three-body disturbing
function of the spacecraft.

It is worthy of note that there is a long interplanetary journey for a spacecraft before the arrival at the target.
In this case, the transformation between $\tau_{s}$ and TCB has exactly the same structure as the Eq.(\ref{TN}), and could be dramatically
simplified as $\tau_{s}-t=-\epsilon^{2}\int(Gm_{\odot}/r_{s\odot}+v^{2}_{s}/2)\mathrm{d}t$ when the probe is far beyond the Hill sphere of any massive body except the Sun.
Therefore, the final $\tau_{s}-t$ during this phase is strongly dependent on the trajectory the spacecraft takes.
However, most of spacecrafts spend their time on this in the quiet mode until crucial orbital maneuvers or scientifically important flybys.
For this reason, we do not take much care of this issue and it is easy to handle indeed.

\normalem
\begin{acknowledgements}
I acknowledge very useful and helpful comments and
suggestions from my anonymous referee. I thank Prof. Cheng HUANG of Shanghai Astronomical
Observatory for his helpful discussions and advice. And I
appreciate the support from the group of Almanac and Astronomical
Reference Systems in the Purple Mountain Observatory of China. This
work is funded by the Natural Science Foundation of China under
Grant Nos. 11103085 and 11178006. This project/publication was made possible through the support
of a grant from the John Templeton Foundation. The opinions expressed in
this publication are those of the authors and do not necessarily reflect
the views of the John Templeton Foundation. The funds from John Templeton
Foundation were awarded in a grant to The University of Chicago which also
managed the program in conjunction with National Astronomical Observatories,
Chinese Academy of Sciences.
\end{acknowledgements}


\label{lastpage}

\end{document}